\documentclass[useAMS,usenatbib]{mn2e}[20pt]
\usepackage{graphicx}
\usepackage[section] {placeins}
\usepackage{subfigure}
\usepackage{float}
\usepackage{color}
\usepackage{amsmath}
\usepackage{url}
\usepackage{natbib}
\def\msun{{\rm\,M_\odot}}

\def\msun{{\rm\,M_\odot}}

\def\h2{${\rm\,H_2}$}

\title[Metal-Poor Damped Lyman Alpha Systems]{The Physical Nature of the Most Metal-Poor Damped Lyman Alpha Systems}

\author[Sihan Yuan$^{1}$ \& Renyue Cen$^{2}$]
{
Sihan Yuan$^{1}$\thanks{E-mail:sihany@princeton.edu}, 
Renyue Cen$^{2}$\thanks{E-mail:cen@astro.princeton.edu}\\
$^{1}$Princeton University, Princeton, NJ 08544\\
$^{2}$Princeton University, Princeton, NJ 08544
}

\begin{document}

\maketitle

\label{firstpage}

\begin{abstract}

Utilizing the high-resolution, large-scale LAOZI cosmological simulations 
we investigate the nature of the metal-poor (${\rm [Z/H]<-2}$) damped Lyman alpha systems (mpDLA) at $z=3$.
The following physical picture of mpDLAs emerges.
The majority of mpDLAs inhabit regions $\ge 20$~kpc 
from the host galaxy center on infalling cold gas streams originating from the intergalactic medium, 
with infall velocity of $\sim 100$ km/s and temperature of $\sim 10^{4}$ K. 
For each host galaxy, on average, about $1\%$ of the area within a radius $150$~kpc is covered by mpDLAs.
The mpDLAs are relatively diffuse ($n_{\rm{gas}} \sim 10^{-2}$ cm$^{-3}$), 
Jeans quasi-stable, and have very low star formation rate ($\dot{\Sigma} \le 10^{-4} \msun \rm{\ yr}^{-1} \rm{\ kpc}^{-2}$). 
As mpDLAs migrate inward to the galaxy center, they mix with high metallicity gas and stellar outflows
in the process, removing themselves from the metal-poor category and rendering 
the central ($\le 5$ kpc) regions of galaxies devoid of mpDLAs. Thus, the central regions of the host galaxies are populated by mostly metal-rich DLAs instead of mpDLAs.
All observables of the simulated mpDLAs are in excellent agreement with observations, 
except the gas density, which is about a factor of ten lower than the value inferred observationally. However, the observationally inferred value is based on simplified assumptions that are not borne out in the simulations.

\end{abstract}


\section{Introduction}

Damped Lyman-alpha Systems (DLAs) are neutral-hydrogen (HI) gas clouds or cloud complexes
with total HI column density, ${\rm N(HI) \geq 10^{20.3}}$~cm$^{-2}$. 
DLAs are fundamentally important because they contain most of the 
neutral gas in the post-reionization universe \citep[e.g.,][]{2000Storrie, 2009prochaska, 2003peroux}.
DLAs provide the reservoir of cold neutral hydrogen 
that is a key link in the fuel supply chain
of star formation, and are thus essential for understanding the formation of galaxies. 
DLAs are most efficiently found through the detection of damped Ly$\alpha$ absorption lines in QSO spectra,
a method pioneered by \citet{1986Wolfe}.
Since then, a concerted effort has led to a significant increase in the sample size of observed DLAs  
\citep[e.g.,][]
{
1991Lanzetta, 
1995Lanzetta, 
2000Storrie, 
2001Ellison, 
2003peroux, 
2004Prochaska, 
2005Prochaska, 
2008Prochaska, 
2009prochaska, 
2009Noterdaeme, 
2009Abazajian, 
2013Jorgenson}. 
Currently there are about one thousand DLAs detected.
For a review see \citet{2005Wolfe}.
The basics of DLAs, including their kinematics, metallicity distribution, sizes and incidence rate,
are reasonably well understood and accounted for in the context of the standard cold dark matter model
\citep[e.g.,][]{2012Cen}.
The physical nature of DLAs evolves with redshift:
At high redshift ($z\ge 2$), DLAs are dominated by dense HI gas at distances from the centers of galaxies
that are about an order of magnitude larger than the radii of contemporary galactic stellar disks.
At low redshift, the HI disks more or less coincide with stellar disks in size become the dominant 
contributor to the DLA population
\citep[e.g.,][]{2012Cen}.

A subset of DLAs, those that are the most metal poor with [Fe/H] $\leq -2.0$, is of special interest.
The most metal-poor DLAs (hereafter mpDLA) are observed to exhibit metallicity as low as [Fe/H]=-3.45 and hence
may contain important information on the first generation of 
Population III stars \citep{2010Ellison,
2011aCooke, 
2011bCooke, 
2012Pettini, 
2014Cooke}. 
These mpDLAs also exhibit an enhancement in their $\alpha$/Fe ratio consistent with 
that seen in the local group dwarf galaxies. 
Thus, these DLAs may also hold a key to understanding 
the formation and history of dwarf galaxies 
\citep[][]{2011bCooke, 2014Cooke}. 
In the last few years, careful surveys have found 
23 mpDLAs with a redshift range of approximately $2 < z < 4.5$ 
\citep[][]{2008Pettini, 2011bCooke, 2013Cooke, 2014Cooke, 2015Cooke}. 
Here in this paper we perform the first, in-depth examination of mpDLAs, 
utilizing the ab initio, high-resolution large-scale LAOZI cosmological hydrodynamic simulations, 
to provide a physical description of mpDLAs and to compare with observationally  
derived physical properties by Cooke et al. 
A particularly important question that we also address in this paper is whether or not mpDLAs originate from
nearly primordial cold gas streams feeding galaxy formation.

The outline of this paper is as follows: \S\ref{sec:methods} covers our simulations and analysis methods.
Specifically, in \S\ref{sec:simulation} we detail our simulations, and explain our method for identifying DLAs and mpDLAs in the simulation in \S\ref{sec:identify}.
Results are presented in \S\ref{sec:results}.
In \S\ref{sec:distribution} we discuss the column density and spatial distribution of our sample of DLAs and mpDLAs. In \S\ref{sec:covfrac} we examine the impact parameter distribution and the covering fraction of DLAs and mpDLAs around galaxies. 
We study the line-of-sight velocity distribution in \S\ref{sec:vlos} and the temperature distribution in \S\ref{sec:Tgas}.
We then examine the gas density distribution of the DLAs and mpDLAs and derive a hydrogen density typical of mpDLAs in \S\ref{sec:ngas}, followed by a discussion on the radial motion of the mpDLAs relative to their host galaxies in \S\ref{sec:vr}. A discussion of the results is offered in \S\ref{sec:discussion} while conclusions are given in \S\ref{sec:conclusion}.

\section{Simulations and Analysis Methods}
\label{sec:methods}

\subsection{The simulation}
\label{sec:simulation}

We make use of the 
{\bf L}arge-scale {\bf A}daptive-mesh-refinement {\bf O}mniscient {\bf Z}oom-{\bf I}n cosmological hydrodynamic simulations,
called {\bf LAOZI Simulations}
to perform an analysis on both the mpDLAs and overall DLAs for comparison.
LAOZI simulations are run using the Adaptive Mesh Refinement Eulerian hydrodynamic code, Enzo \citep[][]{2014Bryan}. 
First we run a low resolution simulation with a periodic box of $120~h^{-1}$ Mpc 
(comoving) on each side and identify a region centered on a cluster of mass of $\sim 3\times 10^{14}\msun$ at $z=0$.
We then resimulate the chosen region embedded
in the outer $120h^{-1}$ Mpc box with high resolution to properly take into account the large-scale tidal field
and appropriate boundary conditions at the surface of a refined region.
The refined region has a comoving size of $21\times 24\times 20h^{-3}$ Mpc$^3$.

A thorough description of the various physical processes considered in the simulations are presented in \citet{2005Cen} and \citet{ 2012Cen}. Here we mention only a few of these processes that are of particular concern for the purpose of this paper. In the simulations, we include a metagalactic UV background \citep[][]{1996Haardt}, and a model for shielding of UV radiation by neutral hydrogen \citep[][]{2005Cen}. Specifically, we employ a local optical depth approximation to mimic self-shielding: each cubic cell is flagged with six HI ``optical depths" on all six faces, and each optical depth is calculated as the product of HI density, HI ionization cross section and scale height, before we compute the mean attentuation (not mean optical depth) of the six values. Similar procedures are carried out for neutral helium and singly ionized helium. This approximation works well, and since DLAs are optically opaque to ionizing photons, we expect any refined treatment of self-shielding is unlikely to have any significant effect. We also include metallicity dependent radiative cooling  in the simulation \citep[][]{1995Cen}. 
Our simulations also solve relevant gas chemistry
chains for molecular hydrogen formation \citep[][]{1997Abel},
molecular formation on dust grains \citep[][]{2009Joung},
and metal cooling extended down to $10~$K \citep[][]{1972Dalgarno}.
We defer some additional discussion of cooling and its relevance to \S\ref{sec:discussion}.

The simulation generates over 3000 galaxies with extremely high spatial resolution ($\leq 111 h^{-1}$ pc physical).
The large galaxy sample size and the high resolution are indispensable to 
study DLAs in detail statistically.
The simulation has the following cosmological parameters that are consistent with
the WMAP7-normalized \citep[][]{2011Komatsu} $\Lambda$CDM model:
$\Omega_M=0.28$, $\Omega_b=0.046$, $\Omega_{\Lambda}=0.72$, $\sigma_8=0.82$,
$H_0=100 h \,{\rm km\, s}^{-1} {\rm Mpc}^{-1} = 70 \,{\rm km\, s}^{-1} {\rm Mpc}^{-1}$ and $n=0.96$.
These parameters are consistent with those from Planck first-year data \citep[][]{2014Planck}
if we average Planck-derived $H_0$ with SN Ia and HST based $H_0$.
For more details, see \citet[][]{2005Cen, 2012Cen, 2014Cen}. 

\citet[][]{2012Cen} compares our simulation to other studies, mostly based on smoothed particle hydrodynamics (SPH) simulations as of the year 2012 \cite[][]{1997aGardner,1997bGardner,
2001Gardner,1998Haehnelt,2004aNagamine,2004bNagamine,
2007Nagamine,2008Pontzen,2009Tescari,2010Hong}. The paper shows that our simulations demonstrate broad agreement with the previous SPH simulations, and the differences between our simulations and the SPH simulations can be attributed to different simulation box sizes and our inclusion of self-shielding in the simulation. 
\citet{2014Bird, 2015Bird} simulated the kinematics, metallicity and distribution of DLAs at $z = 3$ using a series of cosmological simulations utilizing the moving mesh code AREPO \citep[][]{2010Springel}. Their simulation outputs an average DLA metal line velocity width $v_{\rm 90}$ of approximately 100 km/s or 30 km/s, depending on specific model,  consistent with the $v_{\rm 90}$ distribution derived for DLA systems in \citet{2012Cen}. \citet{2014Bird} concludes that DLAs mostly exist in halos of mass $10^{10}-10^{11}h^{-1}$M$_{\odot}$, also consistent with the halo mass range predicted by \citet{2012Cen} (Figure.16). Thus, we can conclude that our simulation is consistent with and complements previous works on DLAs.

\subsection{Identifying DLAs and mpDLAs}
\label{sec:identify}
Since the average redshift of the mpDLAs observed by Cooke et al. is around $z \sim 3$, 
we take a snapshot of the LAOZI simulation box at redshift $z=3$ to compare with the observations. 
We compile a list of the 800 most massive galaxies in the simulation at $z=3$ with stellar mass greater than $6.6\times 10^{9} \msun$. 
For each galaxy we cut out a 300 kpc wide box centered on the galaxy. 
To search for mpDLAs inside each box, we project the neutral hydrogen number density, $n_{\rm{HI}}$, along the direction of line-of-sight (LOS), 
arbitrarily chosen as the $\hat{z}$ direction, 
to obtain the neutral hydrogen column density, ${N_{\rm HI}}$. 
We divide the x-y plane, i.e., the plane perpendicular to the LOS, into 3000-by-3000 pixels, 
producing a total of $9\times 10^6$ LOS columns each of size $100\times 100$ pc$^2$ and depth $300$ kpc.
Then we select all LOS columns containing DLAs as those with neutral hydrogen column density $N_{\rm{HI}} \geq 10^{20.3}$ cm$^{-2}$. 
All statistical results (Figure~\ref{fig:Ndist} to Figure~\ref{fig:Vr-distri}) 
are found using our sample of all LOS columns containing DLAs and mpDLAs among the 800 most massive galaxies.

For each LOS containing a DLA, we compute the $N_{\rm{HI}}$-weighted mean metal density as 
\begin{equation}
\centering
\langle\rho_{Z}\rangle_{\rm{NHI}} \equiv \frac{\int_{\rm{los}} \rho_{Z}n_{\rm{HI}} dl}{\int_{\rm{los}} n_{\rm{HI}} dl},
\label{equ:metaldensity}
\end{equation}
where $\int_{\rm{los}} dl (...)$ is an integral along the LOS. 
Similarly, along each LOS, we calculate the ${N_{\rm HI}}$-weighted mean temperature 
$\langle T\rangle_{\rm{NHI}}$, 
the ${N_{\rm HI}}$-weighted mean radial velocity $\langle v_{\rm{los}}\rangle_{\rm{NHI}}$ 
and the velocity dispersion $\langle \sigma_{\rm{los}}\rangle_{\rm{NHI}}$, among other key physical parameters.
We calculate the $N_{\rm{HI}}$-weighted mean gas density along each LOS, 
$\langle\rho_{\rm{gas}}\rangle_{\rm{NHI}}$. 
We compute the mean DLA metallicity in solar units as
\begin{equation}
\centering
\langle Z\rangle_{\rm{NHI}} \equiv \log{\frac{\langle\rho_{Z}\rangle_{\rm{NHI}}}{0.02\langle\rho_{\rm{gas}}\rangle_{\rm{NHI}}}} 
= \log{\frac{\int_{\rm{los}} \rho_{Z}n_{\rm{HI}} dl}{0.02\int_{\rm{los}} \rho_{\rm{gas}}n_{\rm{HI}} dl}}.
\label{equ:metallicity}
\end{equation}
We select mpDLAs in the simulation as DLAs with $\langle Z\rangle_{\rm{NHI}} \le -2$.

\section{Results}
\label{sec:results}

\subsection{Spatial and column density distributions of mpDLAs}
\label{sec:distribution}

\begin{figure*}
\centering
\includegraphics[width=3.65in]{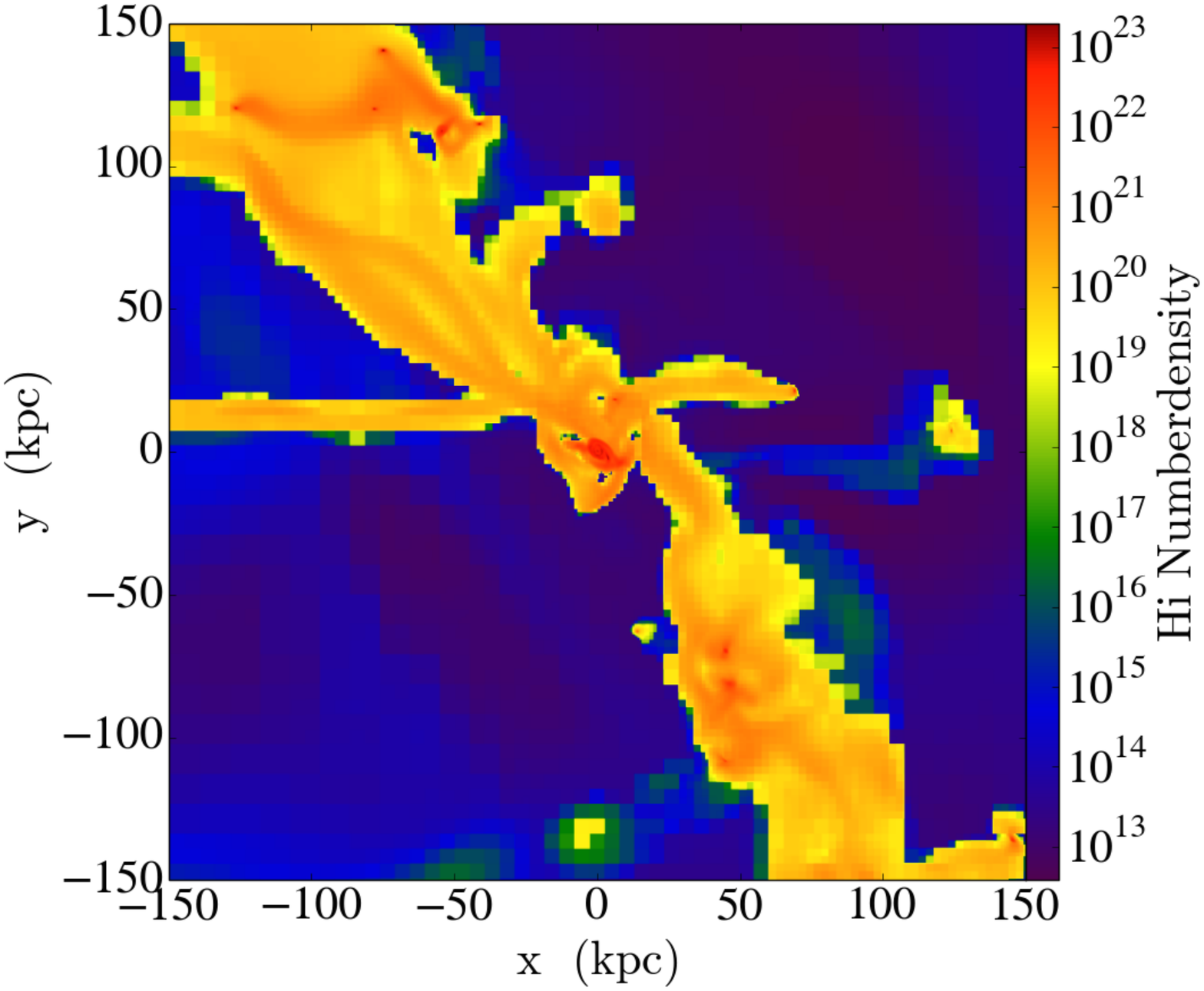}
\hspace{-1.2cm}
\includegraphics[width=3.65in]{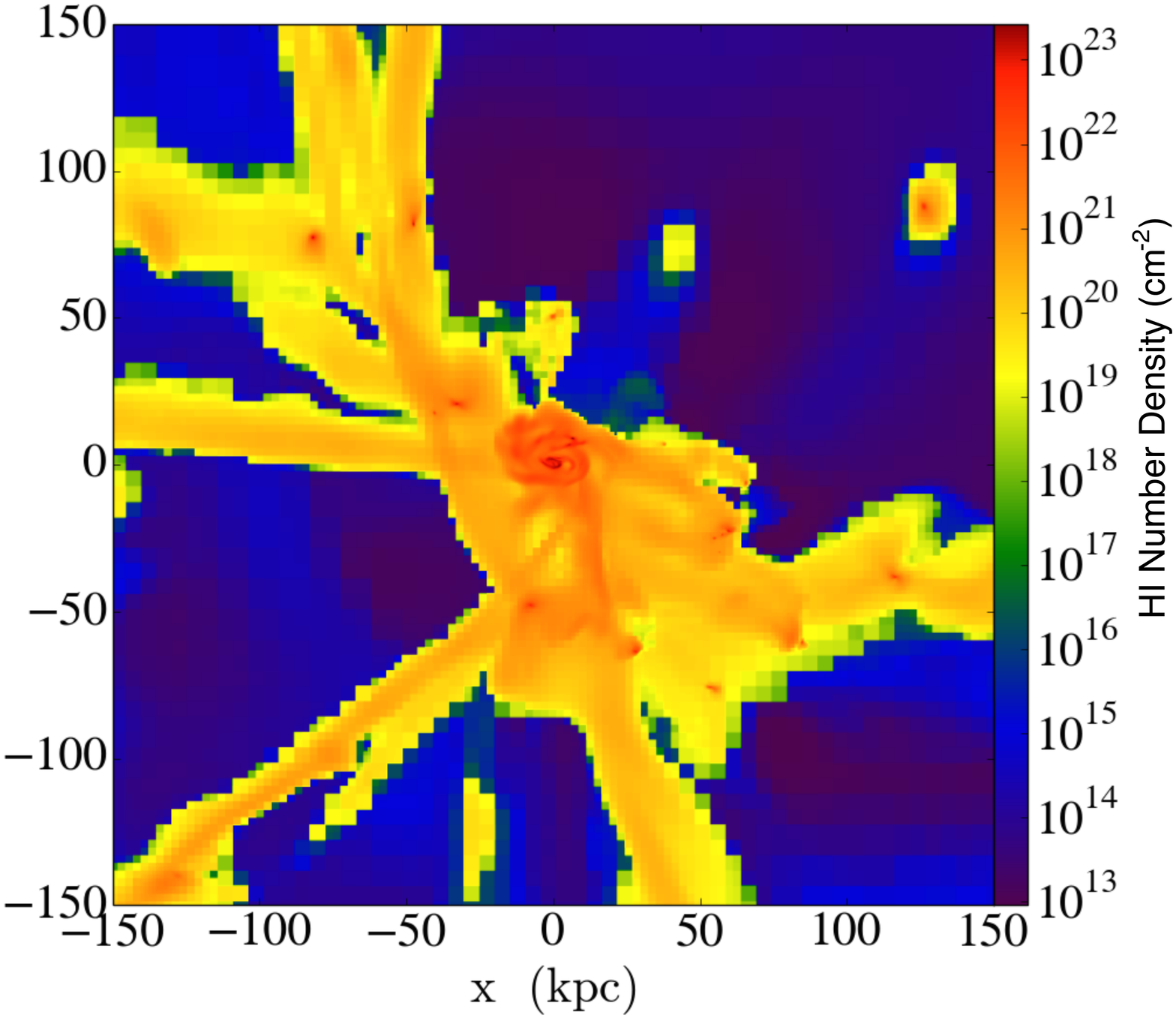}
\includegraphics[width=3.65in]{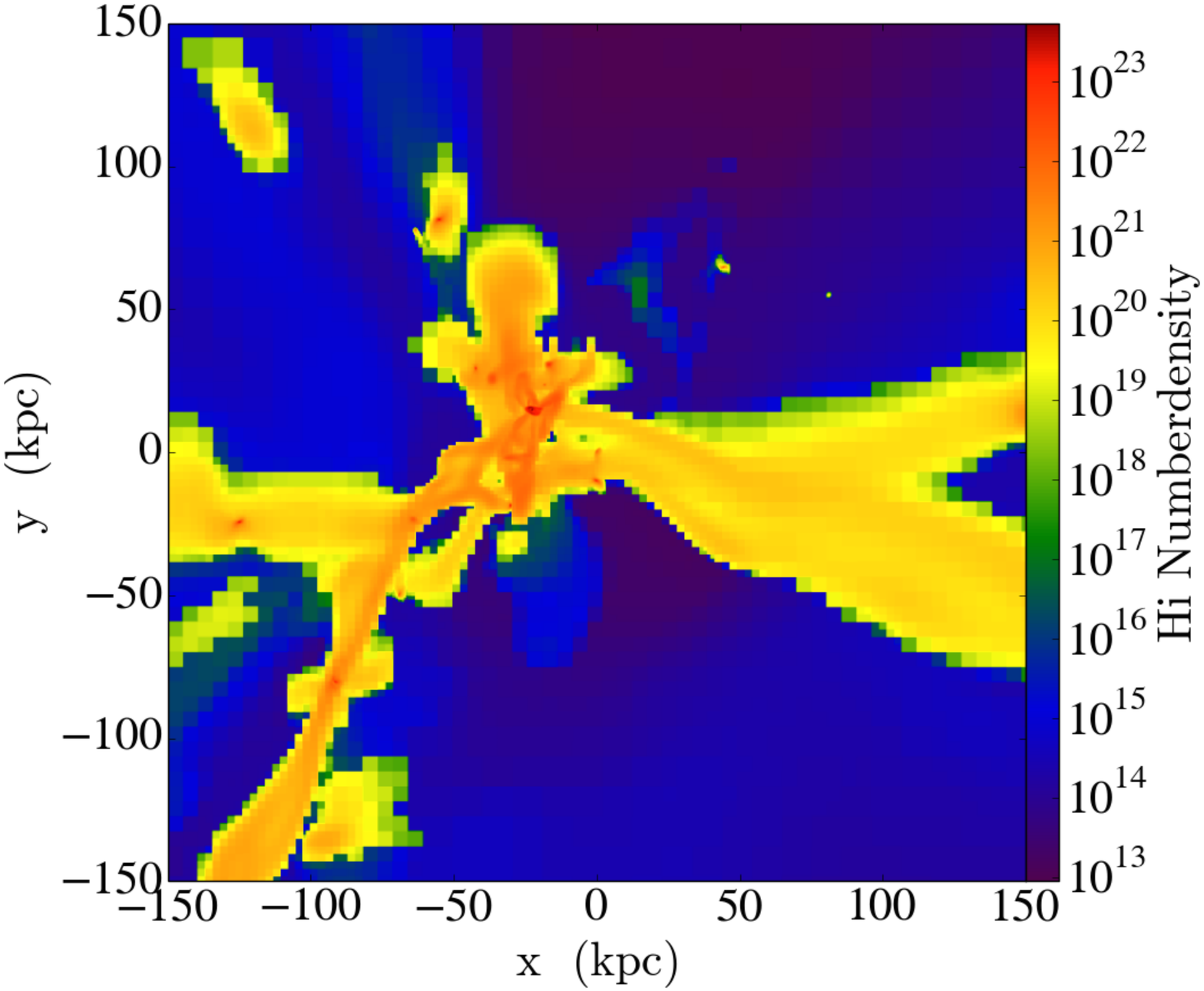}
\hspace{-1.2cm}
\includegraphics[width=3.65in]{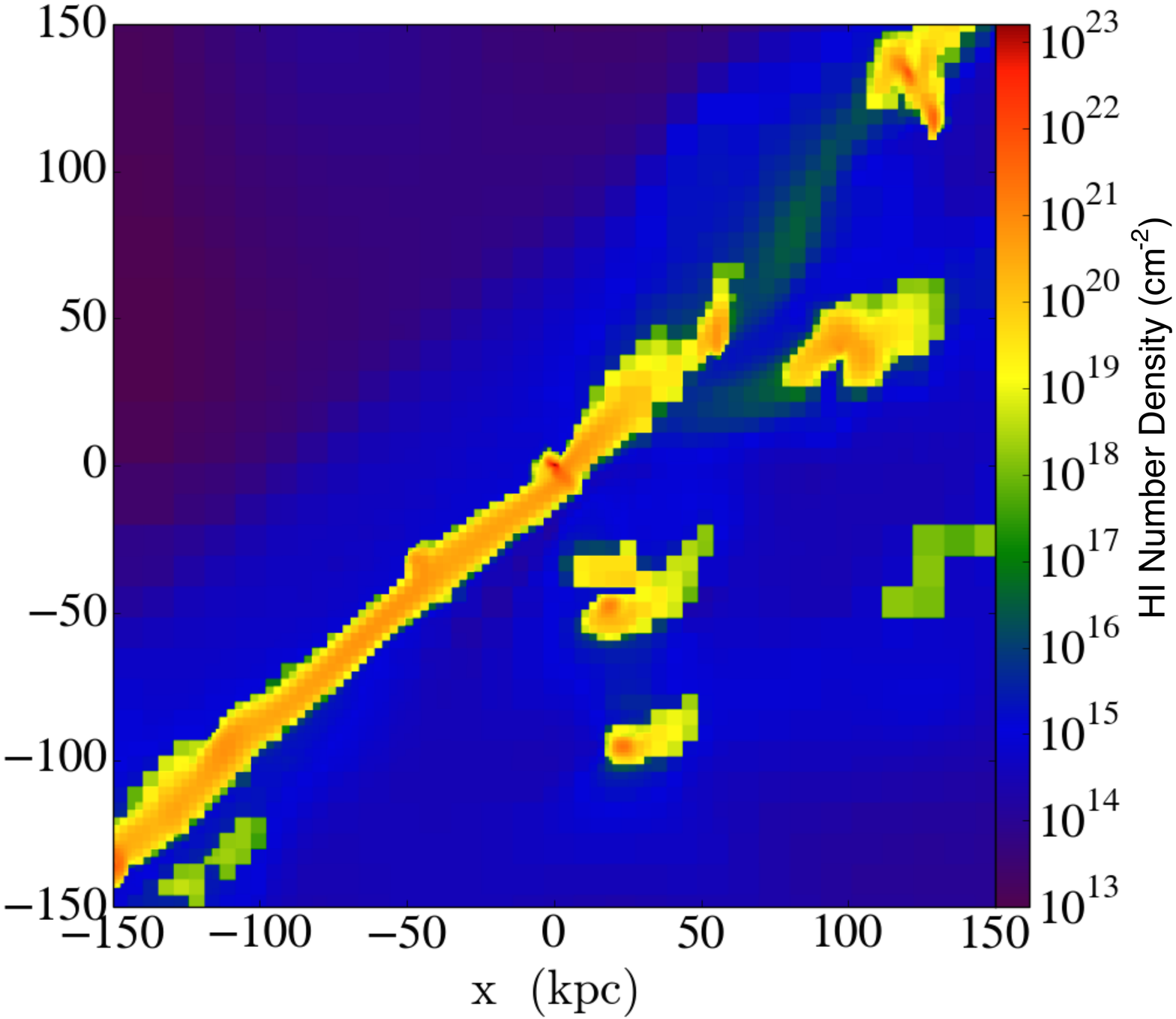}
\caption{shows the neutral hydrogen column density, $N_{\rm HI}$ (cm$^{-2}$), centered 
at four randomly selected galaxies at $z=3$ in our simulation. 
The four galaxies have stellar masses of 
$2.4\times 10^{11} \msun$ (top-left), $1.3\times 10^{11} \msun$ (top-right), 
$2.2\times 10^{10} \msun$ (bottom-left) and $8.3\times 10^{9} \msun$ (bottom-right).}
\label{fig:projs}
\end{figure*}

Before delving into quantitative results,
we visually examine the gas distribution around galaxies.
Figure~\ref{fig:projs} shows neutral density column density ($N_{\rm{HI}}$) projection plots around 
four randomly chosen galaxies.
We see that all four galaxies inhabit intersections of gaseous filaments. 
The inner ``spines" of the filaments have high column densities $N_{\rm{HI}} > 10^{20}$ cm$^{-2}$ to be identified as DLAs.
Recall that the typical stellar disks of galaxies at $z\sim 3$ have sizes of order 1kpc 
\citep[][]{1996bSteidel}.
Thus we can qualitatively conclude that a dominant fraction of the DLAs at high redshift
is due to cold streams that originate from the intergalactic medium,
extending well into the circumgalactic medium and feeding star formation in the inner regions. 
These plots also show that the cold filaments are marginally unstable,
often broken up into islands of neutral hydrogen gas disconnected from the coherent filament structures. 
It is likely that at least some of the cold gas far from the metal rich stellar component of galaxies 
is metal poor and would be habitats for mpDLAs. 
In the subsequent analysis, for uniformity, we consider LOSs within the inscribed circle of each box 
of radius $150$ kpc.  

\begin{figure*}
\centering
\includegraphics[width=4.3in]{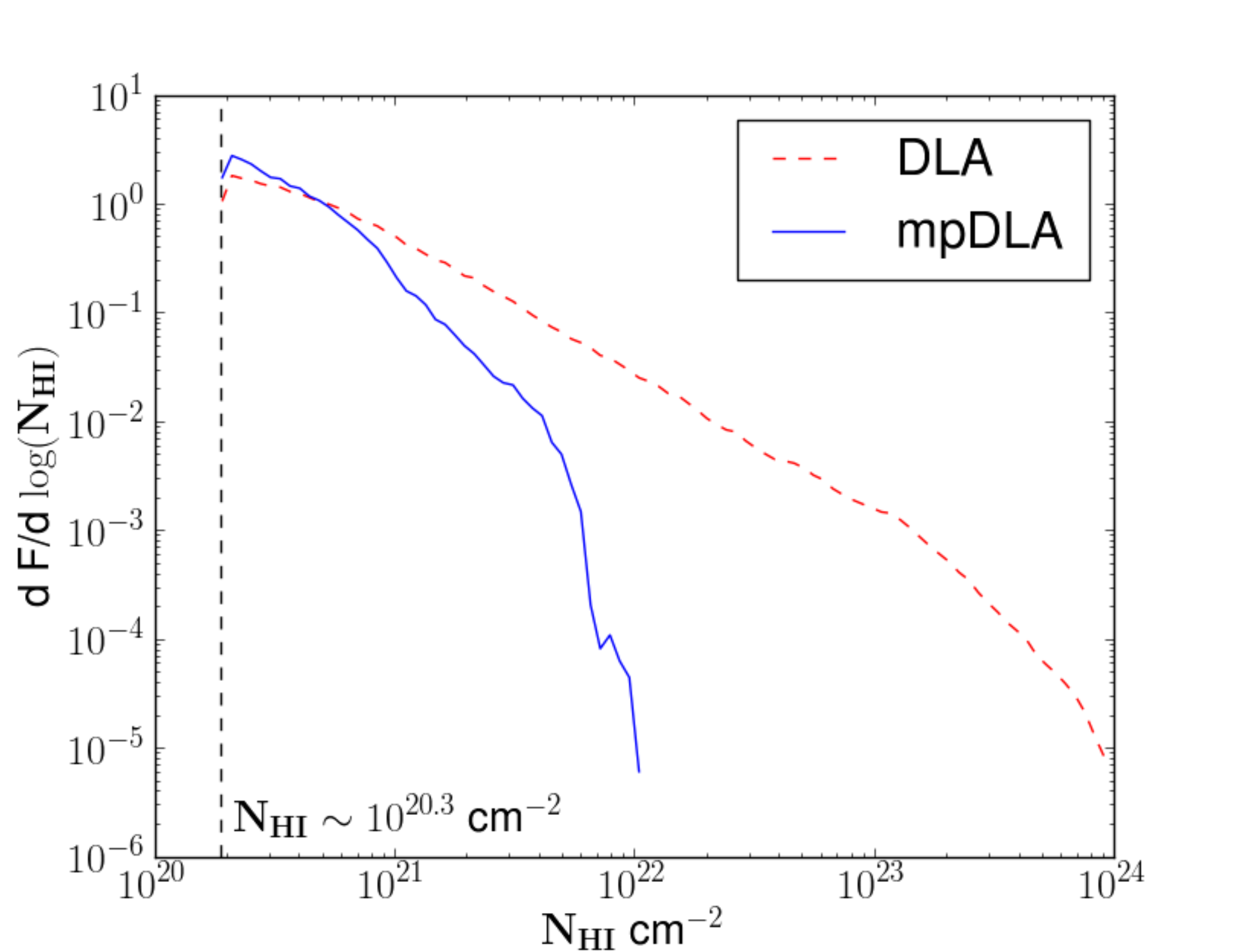}
\caption{shows the probability distribution of the column density $N_{\rm{HI}}$ of 
all DLAs (red dashed curve) and mpDLAs (blue solid curve). 
Note that the section of the DLA curve 
above $N_{\rm{HI}} > 10^{22}$ cm$^{-2}$ is 
physically less meaningful as neutral hydrogen at such density would form molecular hydrogen, 
if proper chemistry were followed.}
\label{fig:Ndist}
\end{figure*}

Figure~\ref{fig:Ndist} shows the column density distribution of all DLAs (red dashed curve) 
and mpDLAs (blue solid curve). 
We see that the mpDLAs tend to have lower 
density and that there is a steep cut-off at around $N_{\rm{HI}} \sim 10^{22}$ cm$^{-2}$, 
whereas the distribution of all DLAs extends into much higher density. 
This suggests that the mpDLAs tend to have more diffuse neutral hydrogen gas structures among all DLAs
and mpDLAs do not directly condense to form molecular gas that has column density exceeding  $10^{22}$ cm$^{-2}$.

\begin{figure*}
\centering
\includegraphics[width=4.3in]{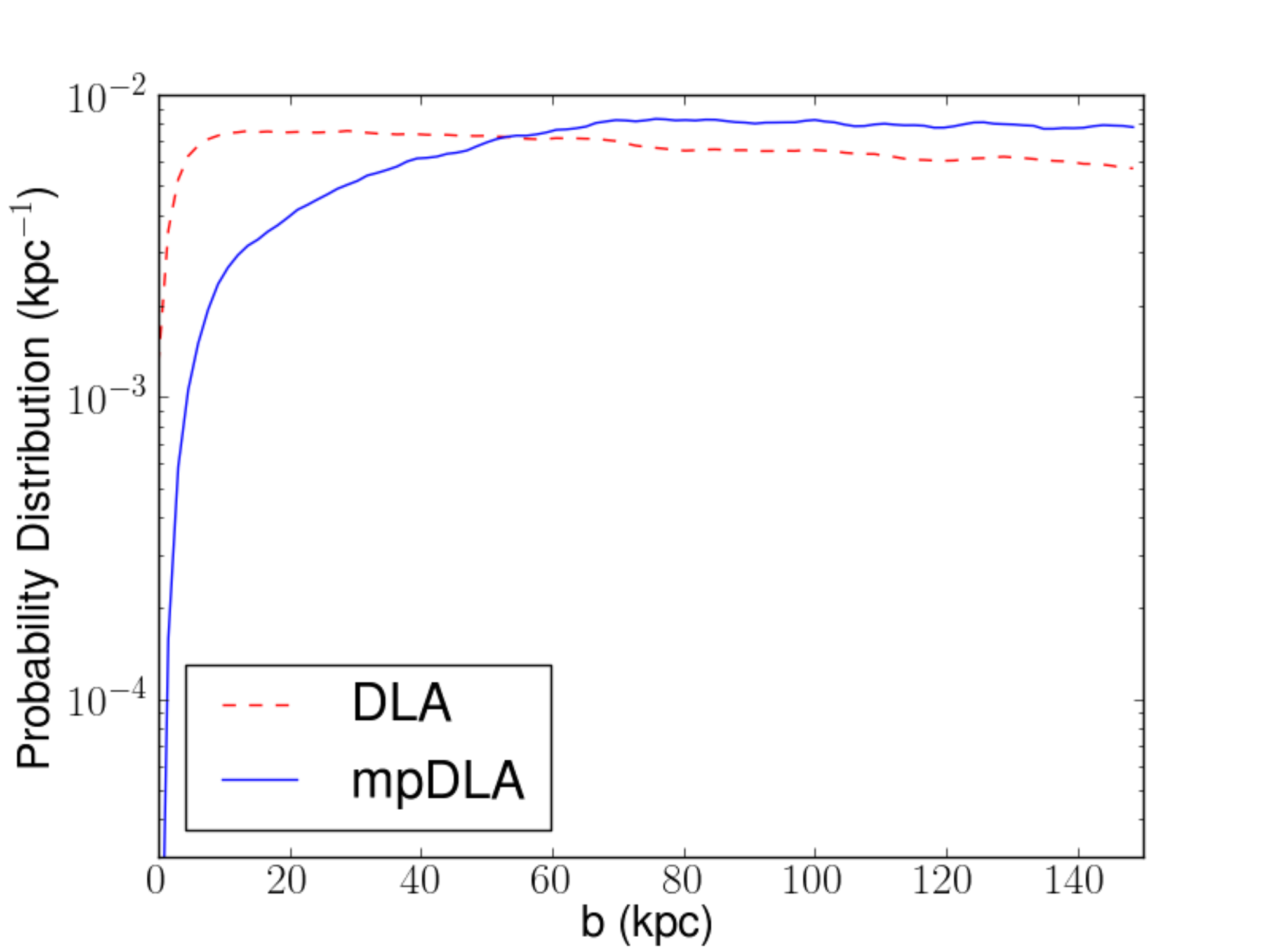}
\caption{shows the probability distribution of the impact parameter $b$ of all DLAs (red dashed curve) 
and mpDLAs (blue solid curve). 
Note that the number of DLAs peaks at around $\sim 10$ kpc,
whereas the mpDLAs tend to inhabit regions more distant ($\ge 50$ kpc) from the host galaxies.}
\label{fig:bdist}
\end{figure*}

\subsection{The impact parameter and covering fraction of mpDLAs}
\label{sec:covfrac}

Figure~\ref{fig:bdist} shows the probability distribution of the impact parameter $b$, defined as the projected distance of the DLA or the mpDLA
to the center of the host galaxy on the plane of the sky. 
The plot indicates that a very small fraction of mpDLAs 
live within the central regions of galaxies with $b \leq 20$ kpc, 
whereas a significant portion of all DLAs inhabit regions as close as $\sim 5$ kpc from the galaxy center.
This is most likely due to unavoidable mixing of metal poor gas with local metal rich gas
at small radii. 
Mixing would hence remove the inflowing initially mpDLA from the metal poor category
upon reaching the central regions.
The median Q$_2$ and interquartile range (Q$_1$, Q$_3$) of the impact parameters for all 
DLAs and mpDLAs are 71 (36, 109) kpc and  87 (56, 118) kpc, respectively. 
The preference for outer regions for mpDLAs is also consistent with their preference for relatively lower densities (compared to the overall DLA population)
seen in Figure~\ref{fig:Ndist}.

\begin{figure*}
\centering
\includegraphics[width=4.3in]{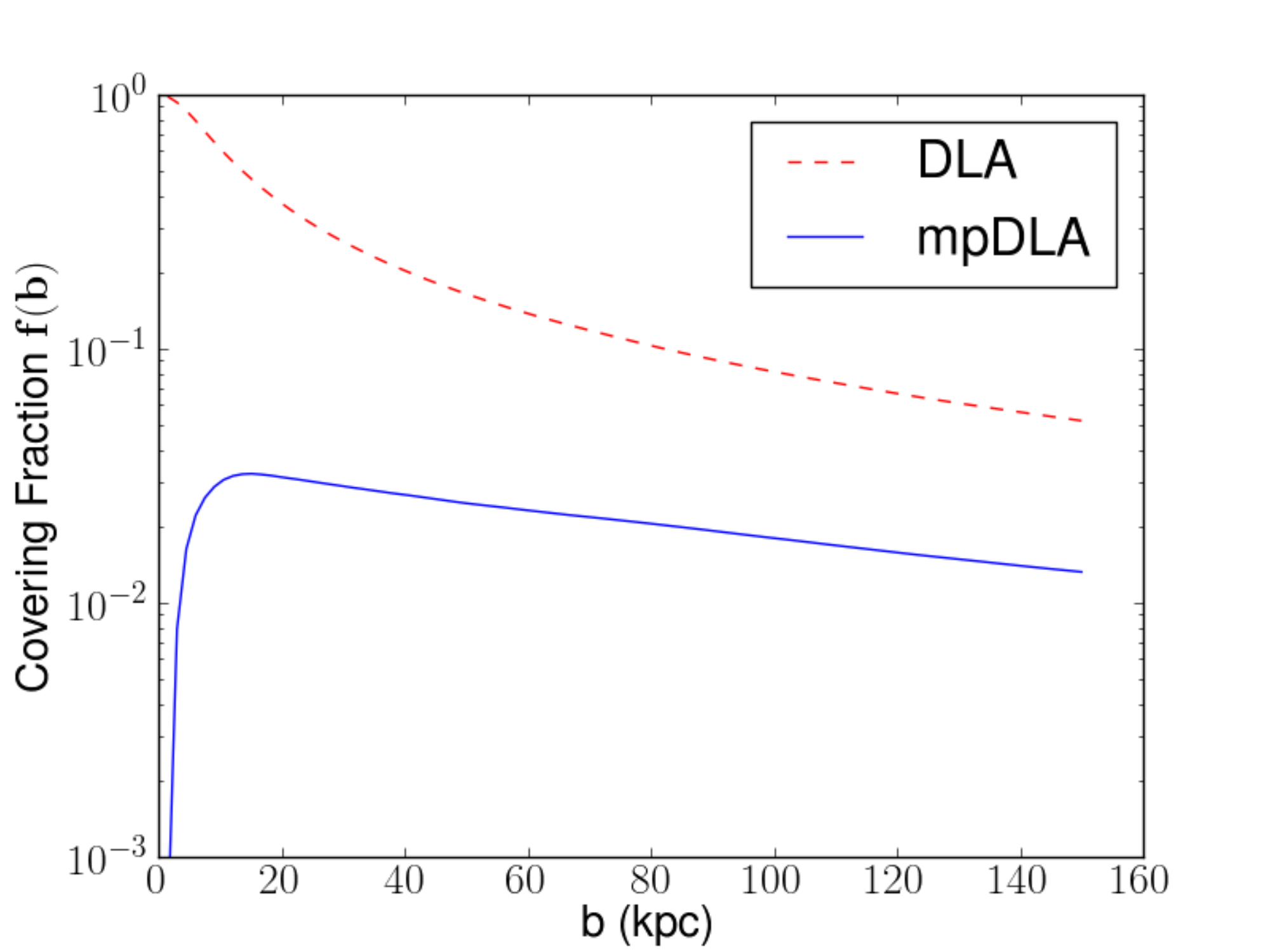}
\caption{shows the average covering fraction of DLAs and mpDLAs as a function of impact parameter $b$. 
The covering fraction is defined as the proportion of area covered by DLAs or mpDLAs within a certain distance of the host galaxies. 
The average is taken over the 800 galaxies in our sample. 
}
\label{fig:avg-covfrac}
\end{figure*}

Figure~\ref{fig:avg-covfrac} shows the average covering fraction of DLAs and mpDLAs as a function of impact parameter $b$.
The covering fraction of DLAs is the fraction of area containing DLAs in the region of interest, and gives us a measure of 
the spatial density of the DLAs. 
Specifically, the covering fraction as a function of impact paramter $b$ is defined as the ratio between 
the area covered by DLAs or mpDLAs and the total area within a radius of $b$ around 
the center of the host galaxy. 
We see that the covering fraction of DLAs starts at $100\%$ at the center of the boxes and drops off 
to $\sim 5.2\%$ toward the edge of the boxes, 
whereas the covering fraction of mpDLAs starts off at $0\%$ and peaks 
at $\sim 3.2\%$ ($b\sim 14$ kpc) before dropping off to $\sim 1.3\%$ towards the edge of the box.
Overall, within $b=150$ kpc, the mpDLAs account for $25\%$ of all DLAs. 
The fraction of mpDLAs decreases with decreasing b, dropping to ($1.8\%,2.5\%,3.0\%,1.8\%$) at $b=(100,50,25,5)$ kpc, respectively. 

This behavior is to be expected because even though the immediate vicinity of the central galaxy is gas rich, it is also metal rich due to star formation. Thus, the immediate vicinity of the centers of galaxies are good candidate sites for DLAs but most mpDLAs live much further from the host galaxies. This is consistent with our picture of mpDLAs as young diffuse clouds of neutral hydrogen far from the galaxy, not yet contaminated by star formation or outflows from the central galaxy.

\subsection{The velocity width of mpDLAs}
\label{sec:vlos}

\begin{figure*}
\centering
\includegraphics[width=4.3in]{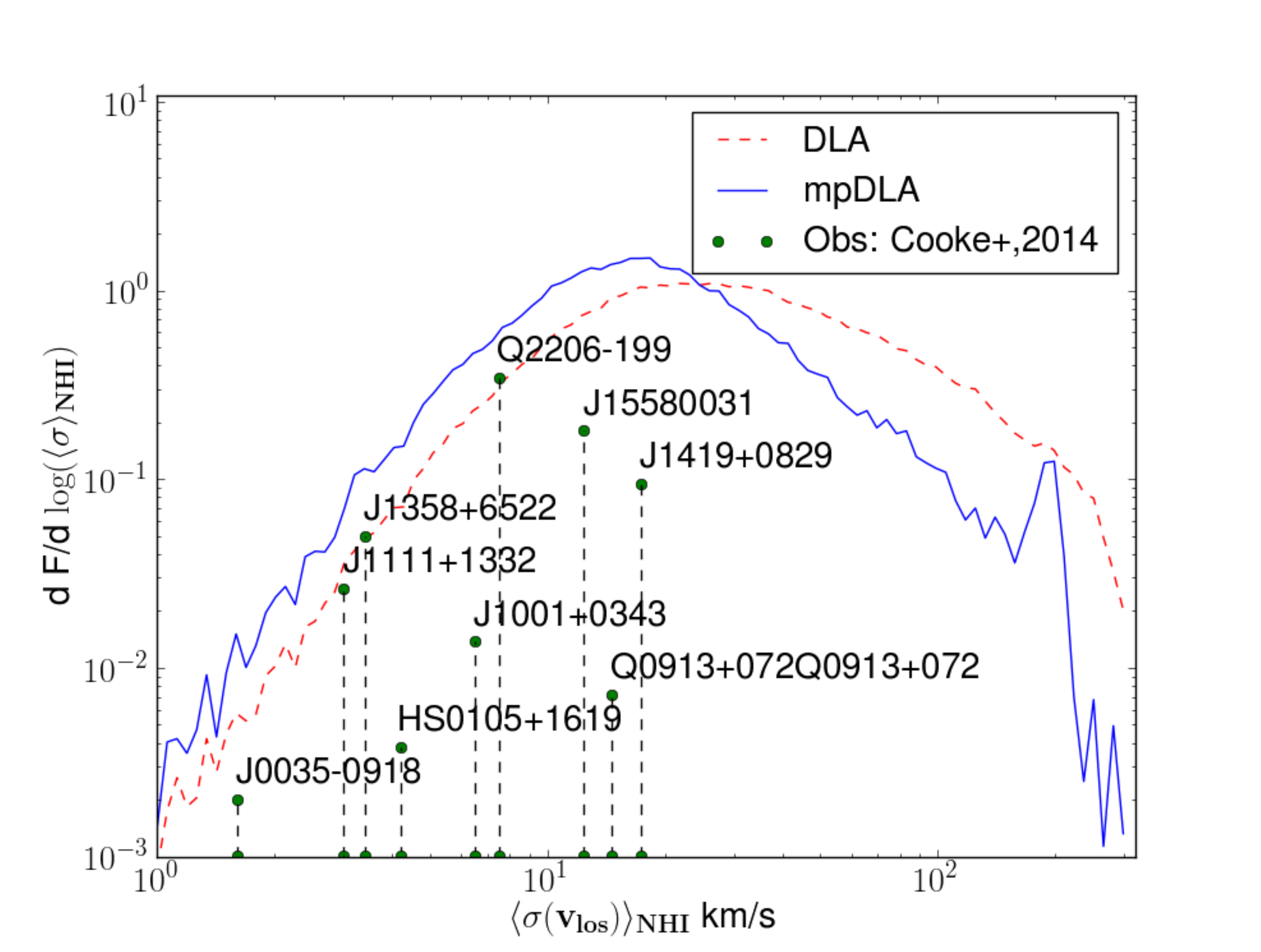}
\caption{shows the probability distribution of the LOS velocity dispersion of all DLAs and mpDLAs. 
The y axis denotes probability per $\log(\langle \sigma(v_{\rm{los}})\rangle_{\rm{NHI}})$. 
The red dashed curve and the blue solid curve correspond to the distribution of 
all DLAs and mpDLAs, respectively. 
Also indicated by vertical dashed lines 
are values of the observed velocity dispersions of a subset of 
\citet[][]{2014Cooke} sample. 
A KS test of the observed data and the simulated data gives a p-value of 0.00402.
}
\label{fig:sigH-distri}
\end{figure*}

To study the kinematic properties of the DLAs and their metal-poor subset, 
we calculate the $N_{\rm{HI}}$-weighted mean LOS velocity dispersion on all DLAs or mpDLAs,
\begin{equation}
\begin{split}
\centering
\langle \sigma_{\rm{los}}\rangle_{\rm{NHI}} & =  \sqrt{\langle v_{\rm{los}}^2\rangle_{\rm{NHI}}-\langle v_{\rm{los}}\rangle_{\rm{NHI}}^2} \\ 
& = \sqrt{\frac{\int_{\rm{los}} v_{\rm{los}}^2n_{\rm{HI}} dl}{\int_{\rm{los}} n_{\rm{HI}} dl} - \left(\frac{\int_{\rm{los}} v_{\rm{los}}n_{\rm{HI}} dl}{\int_{\rm{los}} n_{\rm{HI}} dl}\right)^2}
\end{split}
\label{equ:vlos}
\end{equation}
 Figure~\ref{fig:sigH-distri} shows the probability distributions of $\langle \sigma_{\rm{los}}\rangle_{\rm{NHI}}$ for all DLAs and mpDLAs. 
The median and interquartile range of $\langle \sigma_{\rm{los}}\rangle_{\rm{NHI}}$ are found to be 
28 (16, 51) km/s and 17 (11, 28) km/s for the DLAs and the mpDLAs, respectively. 
The two distributions both peak at around $\sim$20 km/s, although the mpDLAs tend to 
have somewhat lower velocity dispersion compared to the full DLA sample. 
The distribution for all DLAs and that for mpDLAs are similar 
at $\langle \sigma(v_{\rm{los}})\rangle_{\rm{NHI}} < 10$ km/s. 
At $\langle \sigma(v_{\rm{los}})\rangle_{\rm{NHI}} > 10$ km/s, the distribution of mpDLAs 
drops off faster than that of all DLAs. 
A two-sample Kolmogorov-Smirnov (KS) test 
for the velocity dispersion of mpDLAs gives 0.4\%
as the probability of the observed sample and the simulated sample
belonging to the same underlying distribution,
suggesting that the simulated data is inconsistent with observations at $2.9\sigma$ level. 

\begin{figure*}
\centering
\includegraphics[width=4.3in]{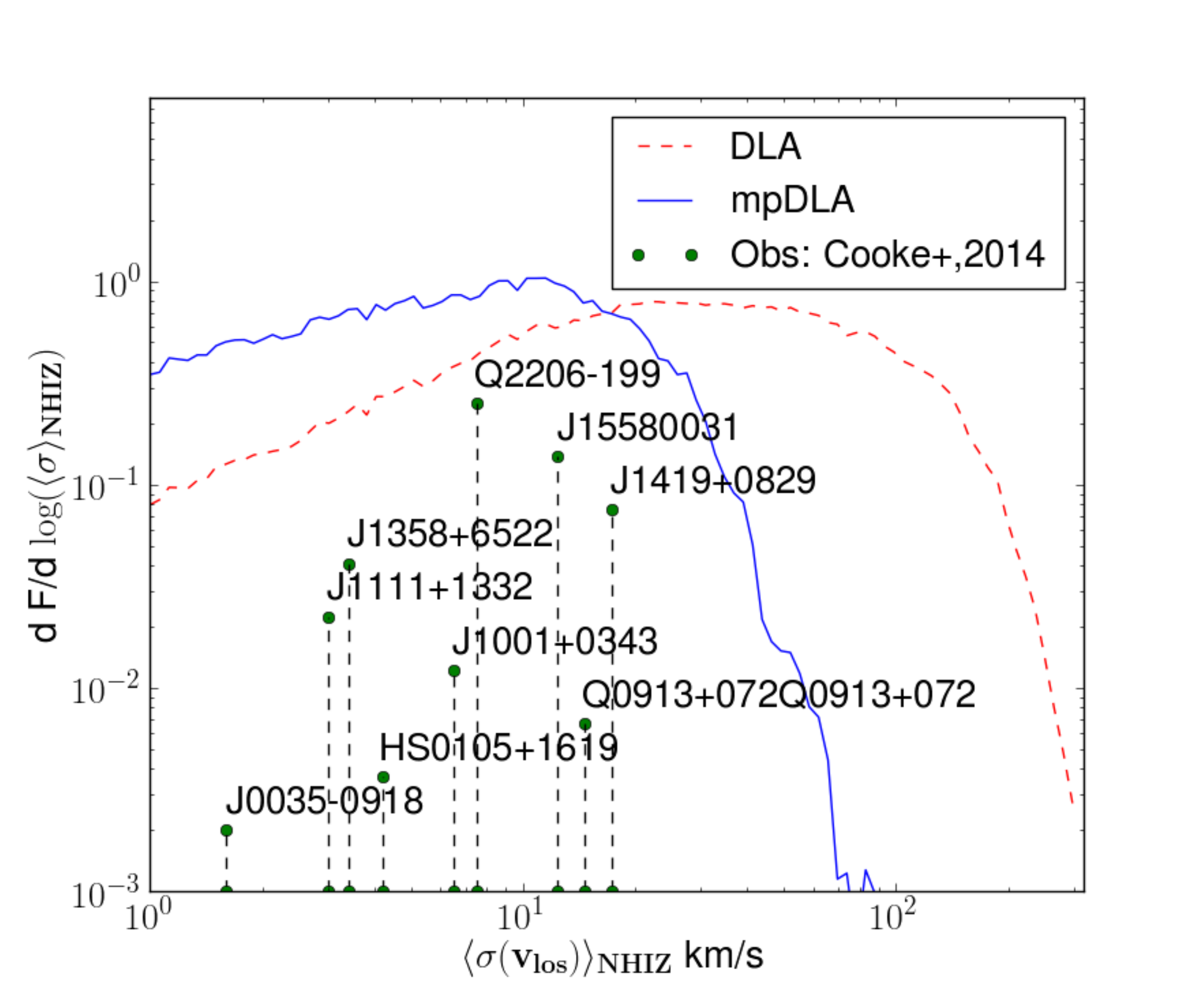}
\caption{shows the probability distribution of the LOS velocity dispersion of all DLAs and  mpDLAs. 
As opposed to Figure~\ref{fig:sigH-distri}, the x axis here denotes the $N_{\rm{HI}}Z$-weighted mean LOS velocity dispersion, 
$\langle \sigma(v_{\rm{los}})\rangle_{\rm{NHIZ}}$. 
The y axis denotes probability per $\log(\langle \sigma(v_{\rm{los}})\rangle_{\rm{NHI}})$. 
The red dashed curve and the blue solid curve represent the distribution of all DLAs and mpDLAs, respectively. 
Also shown as vertical dashed lines are the observed velocity dispersion of a subset of 
\citet[][]{2014Cooke} sample. 
A KS test of the observed data and the simulated data gives a p-value of 0.521.}
\label{fig:sigZ-distri}
\end{figure*}

To further investigate this inconsistency, we note that 
\citet[][]{2014Cooke} derive the 
velocity dispersion from the low-ionization [Fe II] and [Si II] line profiles of the observed mpDLAs. 
In other words, the derived velocity dispersion is representative of the kinematics of the metals inside the DLAs, not necessarily that of the neutral hydrogen. 
Thus, instead of the $N_{\rm{HI}}$-weighting, it is more appropriate to use the $N_{\rm{HI}}Z$ weighting, the product of neutral hydrogen density and metallicity, which traces the total metal density.
We compute the $N_{\rm{HI}}Z$-weighted mean LOS velocity dispersion 
$\langle \sigma_{\rm{los}}\rangle_{\rm{NHIZ}}$ as 
\begin{equation}
\begin{split}
\centering
\langle \sigma(v_{\rm{los}})& \rangle_{\rm{NHIZ}}  = \sqrt{\langle v_{\rm{los}}^2\rangle_{\rm{NHIZ}}-\langle v_{\rm{los}}\rangle_{\rm{NHIZ}}^2} \\
& = \sqrt{\frac{\int_{\rm{los}} v_{\rm{los}}^2n_{\rm{HI}}Z dl}{\int_{\rm{los}} n_{\rm{HI}}Z dl} - \left(\frac{\int_{\rm{los}} v_{\rm{los}}n_{\rm{HI}}Z dl}{\int_{\rm{los}} n_{\rm{HI}}Z dl}\right)^2}
\end{split}
\label{equ:vlosZ}
\end{equation}
\noindent 
The distributions of $\langle \sigma_{\rm{los}}\rangle_{\rm{NHIZ}}$ for all DLA and mpDLAs 
are shown in Figure~\ref{fig:sigZ-distri}. 
We see that the velocity dispersion of all DLAs 
peaks around $30\sim 40$ km/s, while that of the mpDLAs is lower by about $\sim 10$ km/s. 
The median and interquartile range of the distributions are 25 (10, 53) km/s and 6.9 (3.2, 12.7) km/s 
for DLAs and mpDLAs, respectively. 
More significantly, comparing to Figure~\ref{fig:sigH-distri}, 
the distribution of the $N_{\rm{HI}}Z$-weighted mean LOS velocity dispersion 
does not drop off nearly as steeply towards the low end, specifically at $\langle\sigma_{\rm{los}}\rangle < 10$ km/s. 
However, it drops off much faster toward 
higher velocity dispersion, specifically at $\langle\sigma_{\rm{los}}\rangle > 20$ km/s. 
It appears that the $\langle \sigma_{\rm{los}}\rangle_{\rm{NHIZ}}$ distribution for 
mpDLAs is more consistent with the observed velocity dispersions.
Indeed, a KS test of the observed velocity dispersions with the distribution of the $N_{\rm{HI}}Z$-weighted mean LOS 
velocity dispersion yields a p-value of 0.521, 
suggesting consistency between the two data sets. 
This non-trivial success is a very positive outcome, further validating our simulated results. We compare our derived velocity widths for all DLAs to other observational results and simulation results in \S\ref{sec:discussion}. 
We also point out that understanding physically the difference between $\langle \sigma_{\rm{los}}\rangle_{\rm{NHI}}$ and $\langle \sigma_{\rm{los}}\rangle_{\rm{NHIZ}}$ 
would be an interesting topic for future research. 

\subsection{The temperature of mpDLAs}
\label{sec:Tgas}

\begin{figure*}
\centering
\includegraphics[width=4.3in]{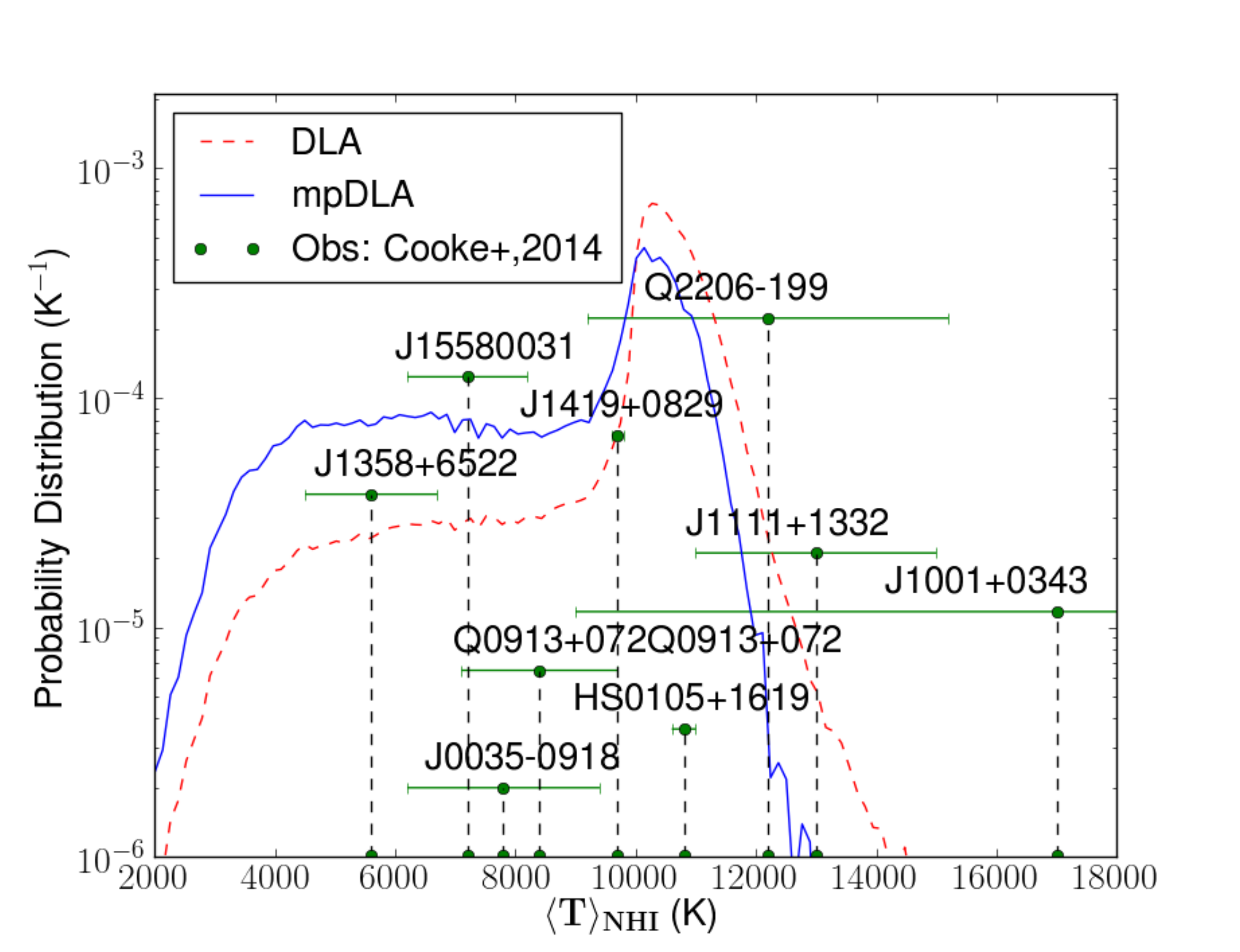}
\caption{shows the probability distribution of the column density $N_{\rm{HI}}$-weighted mean 
temperature of all DLAs (red dashed curve) or mpDLAs (blue solid curve). 
Also shown as vertical dashed lines are the observed temperatures and their errors of a subset of 
\citet[][]{2014Cooke} sample. 
A KS test of the observed data and the simulated data gives a p-value of 0.224.}
\label{fig:T-distri}
\end{figure*}

Figure~\ref{fig:T-distri} shows the probability distribution of the gas temperature of all DLAs and mpDLAs.
The temperature shown is defined as the column density $N_{\rm{HI}}$-weighted mean temperature for all DLAs and mpDLAs, $\langle T\rangle_{\rm{NHI}}$.
We see that the vast majority of DLAs and mpDLAs have temperature in the range
$9.0\times 10^3 \rm{\ K} < \langle \mathit{T}\rangle_{\rm{NHI}} < 1.1\times 10^4 \rm{\ K}$. 
As a whole, the mpDLAs tend to have slightly lower temperature than all DLAs. 
The median and interquartile range of the temperature 
of DLAs and mpDLAs are found to be 1.05 (1.01, 1.09)$\times 10^4$ K and 9.7 (6.6, 10.5)$\times 10^3$ K, respectively. 

\citet[][]{2014Cooke} derive the 
temperature of the observed DLAs by separating the temperature dependent thermal velocity 
dispersion, $\sigma_{\rm{T}}$, from the turbulent velocity dispersion, $\sigma_{\rm{turb}}$. 
The observed total velocity dispersion is given by Equation~\ref{equ:dispersion}, 
\begin{equation}
\sigma_{\rm{obs}}^2 = \sigma_{\rm{turb}}^2 + \sigma_{\rm{T}}^2 = \sigma_{\rm{turb}}^2 + \frac{2 k_{\rm{B}} T_{\rm{gas}}}{m}
\label{equ:dispersion}
\end{equation}
where $T_{\rm{gas}}$ is the temperature of the gas cloud, and $m = m_{\rm{HI}}$ for neutral hydrogen gas cloud. 
Of the 21 mpDLAs in \citet[][]{2014Cooke} sample, 
$\sigma$ decoupling is only possible in 9 systems. 
These derived temperatures are given on Figure~\ref{fig:T-distri}. 
A KS test of the observed temperature and the simulation results gives a p-value of 0.224, 
suggesting that our simulation results are reasonably consistent with observations. 

\subsection{The hydrogen density profile of the mpDLAs}
\label{sec:ngas}

\begin{figure*}
\centering
\includegraphics[width=4.3in]{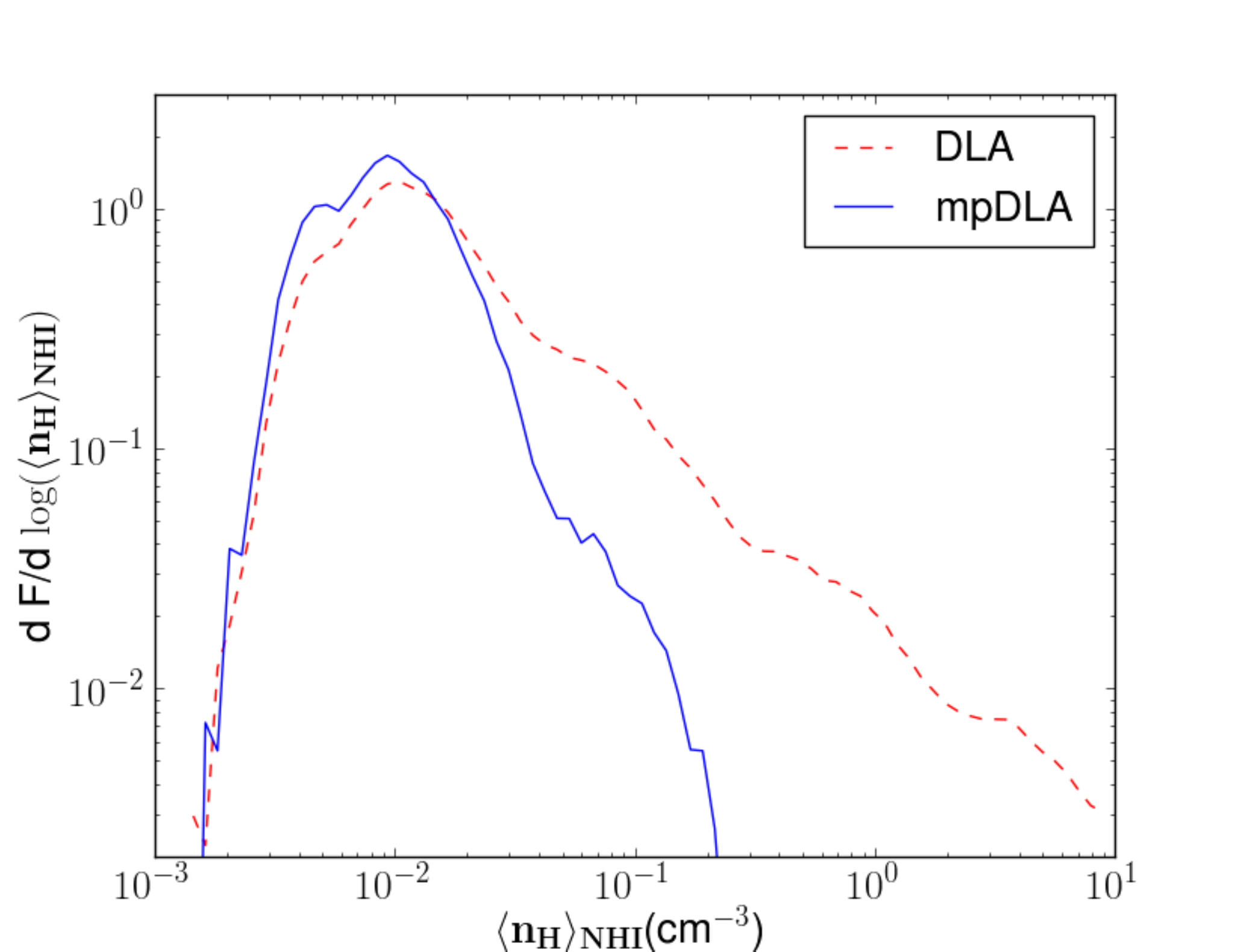}
\caption{shows the probability distribution of the $N_{\rm{HI}}$-weighted mean volumetric neutral hydrogen gas density of 
all DLAs (dashed red) or mpDLAs (solid blue).
}
\label{fig:Ngas-distri}
\end{figure*}

Figure~\ref{fig:Ngas-distri} shows the distribution of neutral hydrogen volume density for both the DLAs and the mpDLAs. 
We compute the mean volume neutral hydrogen gas density of DLAs and mpDLAs 
as $\langle n_{\rm{HI}}\rangle_{\rm{NHI}}$ using the same $N_{\rm{HI}}$-weighting as before,
\begin{equation}
\centering
\langle n_{\rm{HI}}\rangle_{\rm{NHI}} = \frac{\int_{\rm{los}} n_{\rm{HI}}^2 dl}{\int_{\rm{los}} n_{\rm{HI}} dl}.
\label{equ:nHI}
\end{equation}
Both the full DLA sample and the mpDLA sample are dominated, in number, by regions with number density of $\sim 10^{-2}$ cm$^{-3}$. 
Beyond $\sim 2\times 10^{-2}$ cm$^{-3}$ the distribution of all DLAs and that of mpDLAs
deviate from one another, 
with an increasingly smaller fraction of mpDLAs possessing high volumetric density.
The median and interquartile range of the distribution are 0.013 (0.008, 0.024) cm$^{-3}$ and 0.010 (0.006, 0.014) cm$^{-3}$ for DLAs and mpDLAs, respectively. 
\citet[][]{2014Cooke} derive a mean gas density of $n(H)\sim 0.1$ cm$^{-3}$ for mpDLAs, about an order of magnitude higher than our result, we defer a discussion on this discrepancy to Section~\ref{sec:discussion}. 

\subsection{The radial velocities of mpDLAs}
\label{sec:vr}

\begin{figure*}
\centering
\includegraphics[width=4.3in]{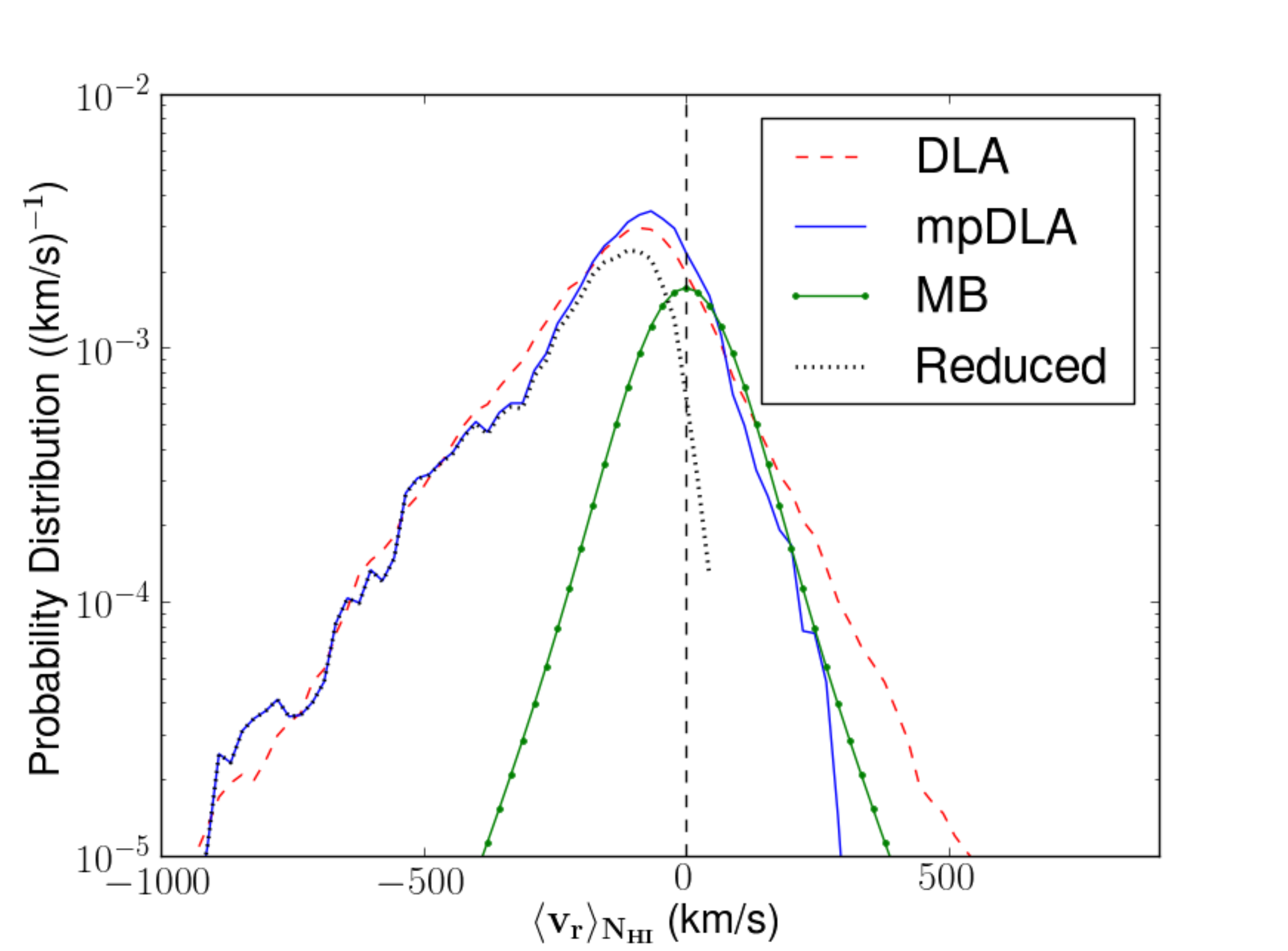}
\caption{shows the probability density distribution of the $N_{\rm{HI}}$-weighted radial velocity of the DLAs 
with respect to the host galaxies. 
A positive velocity corresponds to outflow and a negative velocity corresponds to inflow. 
The red dashed curve and the blue solid curve are the normalized probability distribution of 
all DLAs and mpDLAs, respectively. The green curve with dots is the expected radial velocity distribution 
if the velocity distribution is a Maxwell-Boltzmann (MB) distribution. 
The black dotted curve is the resulting radial velocity distribution after subtracting the MB distribution from the radial velocity distribution of the mpDLAs, 
with the MB curve (green curve with dots) normalized to have the same integral probability
as the mpDLA curve (blue solid curve) at $\langle v_{\rm{r}}\rangle_{\rm{NHI}}>0$.}
\label{fig:Vr-distri}
\end{figure*}

Another important aspect of the kinematics of mpDLAs is their radial velocity relative to their host galaxies. 
Figure~\ref{fig:Vr-distri} shows the probability distributions of the $N_{\rm{HI}}$-weighted average radial velocity 
$\langle v_{\rm r}\rangle_{\rm{NHI}}$ of all DLAs and mpDLAs, defined as 
\begin{equation}
\centering
\langle v_{\rm r}\rangle_{\rm NHI} \equiv \frac{\int_{\rm{los}} v_{\rm r} n_{\rm HI} dl}{\int_{\rm{los}} n_{\rm{HI}} dl}.
\label{equ:vr}
\end{equation}
A word of clarification is useful for $\langle v_{\rm r}\rangle_{\rm NHI}$,
which is the mean of the magnitude of the radial velocity $v_{\rm r}$ of each gas element with sign:
A positive $v_{\rm r}$ corresponds to outflow and a negative one corresponds to inflow. 
The red dashed curve and blue solid 
curve show the normalized probability distributions of the DLAs and the mpDLAs, respectively. 
Since we are most interested in the nature of inflow or outflow of the DLAs,
we construct a distribution model for the random motion of mpDLAs, which will then be subtracted 
from the overall radial velocity distribution, as follows.

First, we construct the average radial velocity distribution of the mpDLAs assuming the 
motion is random and is described by Maxwell-Boltzmann (MB) distribution. 
Each galaxy has a characteristic velocity dispersion as a function of its dark matter halo mass 
$\sigma(M)$ \citep[e.g.,][]{1997Navarro}. 
Using the characteristic velocity dispersion, 
we construct a random radial velocity distribution for each galaxy based on the MB
distribution, $P_{\rm{MB}}(\sigma(M))$. 
Then, we compute the radial velocity distribution of the 
mpDLAs, $P_{\rm{MB}}(\langle v_{\rm{r}}\rangle_{\rm{NHI}})$, by averaging the random 
radial velocity distribution of each galaxy weighted by the mpDLA 
covering fraction of each galaxy, $f_{\rm{mpDLA}}$, as
\begin{equation}
P_{\rm{MB}}(\langle v_{\rm{r}}\rangle_{\rm{NHI}}) = C\sum_i P_{\rm{MB}}(\sigma(M_i))\times f_{\rm{mpDLA},i}
\label{equ:Vr-distri-MB}
\end{equation}
with $C$ being a normalization constant. 
The sum in our case is over the 800 galaxies in our catalog, 
and we exclude from the sample satellite galaxies (identified as galaxies within the virial radius of a more massive galaxies)
because of ambiguity of defining halo mass in this case. 
The resulting radial velocity distribution is shown as the green curve with dots
in Figure~\ref{fig:Vr-distri}. 
Note that the normalization constant $C$ is chosen so that 
the area under the positive half of the MB curve is the same as that 
of the positive part of the mpDLA curve (blue solid curve).
Then we subtract this MB distribution from the overall 
radial velocity distribution of the mpDLAs to obtain the 
coherent (inflow or outflow) part of the radial velocity distribution of the mpDLAs, 
shown as the dotted black curve in Figure~\ref{fig:Vr-distri}. 
It is evident that almost all mpDLAs are in-falling toward their host galaxies, suggesting that almost all mpDLAs 
are part of the in-falling neutral gas streams along the cold filaments. 

To summarize the numerical results,
Table~\ref{tab:avgs} 
lists the median values and interquartile ranges of the various physical properties of 
the mpDLAs in our simulated results that we have investigated. 
The following physical picture for mpDLAs emerges:
the majority of mpDLAs (${\rm [Z/H]<-2}$) 
are young (i.e., just accreted), relatively diffuse neutral hydrogen gas structures with 
$n_{\rm{gas}} \sim 10^{-2}$ cm$^{-3}$. 
They inhabit mostly regions $>20$ kpc from the host galaxy center on infalling cold gas streams originating from the intergalactic medium, 
with an average infall velocity of $\sim 100$ km/s, an average gas temperature of $\sim 10^{4}$ K
and $\sim 1\%$ covering fraction within a radius of $150$ kpc of high-z galaxies. 
They are expected to continue to migrate inward to the galaxy center while being mixed with high metallicity gas and stellar outflows
in the process, removing themselves from the metal-poor category and rendering 
the central ($\le 5$ kpc) regions of galaxies devoid of mpDLAs.

\section{Discussion}
\label{sec:discussion}

\begin{table*}
\caption{Median Q$_2$ and interquartile range (Q$_1$, Q$_3$) of key physical results}
\setlength{\tabcolsep}{20pt}

\begin{tabular}{lll}
\hline\hline
{Quantity} & {All DLAs Q$_2$ (Q$_1$, Q$_3$)} & {Metal-poor DLAs Q$_2$ (Q$_1$, Q$_3$)} \\\hline \vspace{0.1cm}
$\langle b\rangle_{\rm{NHI}}$ (kpc) & 71 (36, 109) &  87 (56, 118)\\
$\langle T\rangle_{\rm{NHI}}$ (K) & 1.05 (1.01, 1.09)$\times 10^4$ & 9.7 (6.6, 10.5)$\times 10^3$ \\
$\langle n_{\rm{gas}}\rangle_{\rm{NHI}}$ (cm$^{-3}$) & 0.013 (0.008, 0.024) & 0.010 (0.006, 0.014) \\
$\langle \sigma_{\rm{los}}\rangle_{\rm{NHI}}$ (km/s) & 28 (16, 51) & 17 (11, 28) \\
$\langle \sigma_{\rm{los}}\rangle_{\rm{NHIZ}}$ (km/s) & 25 (10, 53) & 6.9 (3.2, 12.7) \\
$\langle v_{\rm{r}}\rangle_{\rm{NHI}}$ (km/s) & $-102$ ($-214$, $-14$) & $-90$ ($-189$, $-15$) \\\hline
\end{tabular}
\label{tab:avgs}
\end{table*}

One potential problem with our calculation of impact parameter in \S\ref{sec:covfrac} is that we do not know for sure if every DLA and mpDLA inside a certain box is associated with the galaxy at the center of the box. One can imagine cases (1) where the DLA is associated gravitationally with a companion galaxy also inside the box but closer to the DLA under consideration, and (2) where the DLA is gravitationally bound to a galaxy outside the 300kpc box. The impact parameter for Case (1) is still physically meaningful since the companion galaxy the DLA is bound to is often also gravitationally bound to the galaxy at the center of box. Thus, we can consider these DLAs to be bound to the central galaxy in a generalized sense. The impact parameter for Case (2) is less meaningful, but we can estimate the contribution of Case (2) DLAs to our sample. Specifically, if we examine Figure~\ref{fig:Vr-distri}, we see that a very small fraction of mpDLAs exhibit positive radial velocity, i.e. there is very little outflow. Since one expects Case (2) DLAs to be more likely to move away from the central galaxy and toward the galaxy that it is bound to, we can infer from Figure~\ref{fig:Vr-distri} that this case is not common. Thus, we conclude that the majority of our selected mpDLAs are indeed gravitationally bound to the galaxy at the center of each box. Thus, our calculation of impact parameter still conveys the correct meaning. 

One key result of our simulations is the metal-weighted LOS velocity dispersions of DLAs and mpDLAs, as derived in \S\ref{sec:vlos}. Observationally, the Lyman-$\alpha$ line is saturated for DLAs, so the kinematics of the DLAs is derived from the metal lines such as Si II, which traces HI in DLAs. Such metal lines are measured using an important statistic called $v_{\rm 90}$, defined as the velocity width that covers 90$\%$ of the total optical depth of the observer \citep[][]{1997Prochaska}.
\citet{1997Prochaska} derives the $v_{\rm 90}$ statistic of DLAs from observation and demonstrate a peak at around 70 km/s. \citet{2015Bird} simulates DLAs with the AREPO code and demostrates a $v_{\rm 90}$ peak at 30 km/s or 100km/s depending on the model. 
$v_{\rm 90}$ is related to our velocity dispersion along line of sight through,
\begin{equation}
\centering
v_{\rm 90} \cong 3.3 \langle \sigma_{\rm{los}}\rangle_{\rm{NHIZ}}
\end{equation}
In \S\ref{sec:vlos}, we derived the median metal weighted velocity dispersion $\langle \sigma_{\rm{los}}\rangle_{\rm{NHIZ}} \cong 25$ km/s, which corresponds to $v_{\rm 90} \cong 82$ km/s. This result is consistent with both \citet{1997Prochaska} and \citet{2015Bird}, confirming the credibility of our simulations and lending support to the DEF, FAST, SMALL models in \citet{2015Bird}. 

Another key result of our simulations is the average temperature of the mpDLAs, as derived in \S\ref{sec:Tgas}. To ensure the validity of this result, a proper treatment of heating and cooling in our simulation is critical. Notice that our resolution of $\leq 111 h^{-1}$ pc physical is not enough to capture the detailed chemistry within the densest gas clouds, primarily molecular clouds, 
which are approximately 100 pc in size. However, DLAs are much larger systems ($\sim 10$ kpc), which means that the resolution of our simulation is more than enough to capture the detailed dynamics of the systems. Another point is that there is little molecular cloud formation or star formation within DLAs 
\citep[e.g.,][]{2012Cen, 2010Fumagalli, 2015Fumagalli}, 
allowing for adequate results to be obtained with moderate resolution as in our simulations. 
It is possible that small granules of star formation and molecular cloud formation do exist in our DLAs, but their effect on the overall physical properties of the DLAs are insignificant. Thus, we conclude that our simulations are well resolved for our purpose, and our derivation of the average temperature of DLAs and mpDLAs is appropriate.

In \S\ref{sec:ngas}, we derived a median neutral hydrogen density of 0.013 cm$^{-3}$ and 0.010 cm$^{-3}$ for DLAs and mpDLAs respectively, approximately an order of magnitude lower than the value derived by \citet[][]{2014Cooke}. 
To understand this apparent discrepancy, 
we examine the procedure used by 
\citet[][]{2014Cooke} in deriving the mean gas density of mpDLAs.
They conduct a \textit{Cloudy} simulation of a mpDLA irradiated by a  UV background described by 
\citet[][]{2012Haardt}, which yields a relationship between 
ion ratio [Si III]/[Si II] and the number density of neutral hydrogen. 
Then the number density can be inferred given the observed metal absorption profiles of mpDLAs. 
The \textit{Cloudy} simulation assumes a DLA as a constant density slab with a prescribed constant metallicity. 
We note that these assumptions are not universally consistent with the results of our simulations.
Therefore, we think that the discrepancy could be partly the result of the oversimplifications in the modeling adopted by \citet[][]{2014Cooke}.


One way of constraining the hydrogen density of the mpDLAs is by considering the star formation rate of these systems. 
\citet{2012Krumholz} propose the following formula for surface star formation rate $\dot{\Sigma}$, 
\begin{equation}
\dot{\Sigma} = \epsilon_{\rm{ff}}\frac{\Sigma}{t_{\rm{ff}}}
\label{equ:sfr}
\end{equation}
where $\epsilon_{\rm{ff}}$ is a dimensionless factor with a weak dependence on other quantities, generally approximately as $\epsilon_{\rm{ff}}\approx 0.01$ \citep{2005Krumholz}. 
$\Sigma$ is the column density of neutral hydrogen, and can be taken from observation.
The free-fall time is related to the hydrogen density by $t_{\rm{ff}}=\sqrt{3\pi/32Gm_{\rm{p}}n_{\rm{HI}}}$, 
where $m_{\rm p}$ is the proton mass and $G$ is the gravitational constant.
Thus, if we can derive a star formation rate of mpDLAs from observations, 
we will be able to place a constraint on the density of hydrogen in these systems. 

\citet{2015Fumagalli})
observe 32 DLAs at redshift $z \sim 2.7$ and suggest an upper bound on the star formation 
rate in DLAs, $\dot{\Sigma}_{\rm{obs}} < 10^{-2.6}-10^{-1.5} \msun \rm{\ yr}^{-1} \rm{\ kpc}^{-2}$ \citep{2010Fumagalli, 2014Fumagalli, 2015Fumagalli}. 
A few of the observed DLAs have confirmed galaxy associations and exhibit column density 
close to $N_{\rm{HI}}\sim 10^{20.3}$ cm$^{-2}$, specifically, 
$N_{\rm{HI}} = 10^{20.4}$ cm$^{-2}$, $10^{20.6}$ cm$^{-2}$ and $10^{20.0}$ cm$^{-2}$ \citep{2010Fumagalli}.
Utilizing these observational data, our derived $n_{\rm{HI}}\approx 0.01$ cm$^{-3}$ and Equation~\ref{equ:sfr}, 
we can derive 
the estimated star formation rate of our DLA samples, $\dot{\Sigma} \approx 10^{-4.4} \msun \rm{\ yr}^{-1} \rm{\ kpc}^{-2}$, 
which is consistent with the upper bound set by \citet{2015Fumagalli}.

Independently, knowing the characteristic density of the mpDLAs, we can calculate their Jeans length. Thus, using $n_{\rm{gas}}=10^{-2}$ cm$^{-3}$ and $T=10^4$ K, we have, 
\begin{equation}
\lambda_J = \sqrt{\frac{15 k_B T}{4\pi G m_{\rm{HI}}^2 n_{\rm{gas}}}} \approx 10 \rm{\ kpc}.
\end{equation}
Note that the product of $10$ kpc and the characteristic density $n_{\rm{gas}}=10^{-2}$ cm$^{-3}$ is about $3\times 10^{20}$ cm$^{-2}$,
fully consistent with the column density of DLAs.
The earlier projection plots in Figure~\ref{fig:projs} show that
if we identify the mpDLAs with the inter-galactic filaments,
their sizes are consistent with about 10kpc, or the Jeans length. 
This suggests that the typical DLAs or mpDLAs in our simulations 
are at the boundary of Jeans instability.
This is consistent with our picture of mpDLAs as newly formed dense neutral hydrogen structures. 
They are relatively diffuse (compared to more dense DLAs and star forming regions) and quasi-stable, 
with star formation barely starting to take place ($\dot{\Sigma} \le 10^{-4} \msun \rm{\ yr}^{-1} \rm{\ kpc}^{-2}$).  

One general concern we want to address is that we placed a mass cut on the host galaxies of our sample of DLAs and mpDLAs by only considering the 800 most massive host galaxies in the simulation box. This cut is applied to ensure that
our simulated galaxies considered are resolved beyond any doubt.
Here we test the convergence of our key physical results (those presented in Table~\ref{tab:avgs}) by calculating them for various host galaxy mass cuts. As two examples, Table~\ref{tab:converg} shows the median gas temperature $\langle T\rangle_{\rm{NHI}}$ and median gas density $\langle T\rangle_{\rm{NHI}}$ of all DLAs calculated for five mass cuts. The mass cuts $M_{\rm star}^{\rm cut}$ are given in terms of the stellar mass of the host galaxies. The five mass cuts correspond to the stellar mass of the 400th, 500th, 600th, 700th and the 800th most massive host galaxy in the simulation. We see that the median gas temperature varies by $\sim 0.03\%$ and the gas density varies by $\sim 1\%$ as we reduce the mass cut by half. Similar results hold for all other quantities presented in Table~\ref{tab:avgs} and for mpDLAs as well. Thus, it is safe to conclude that key physical parameters of DLAs and mpDLAs have converged, and the arbitrarily chosen mass cut has minimal effect on our results. 

\begin{table}
\caption{Sample key results calculated at different mass cuts}
\setlength{\tabcolsep}{13pt}

\begin{tabular}{lll}
\hline\hline
{$M_{\rm star}^{\rm cut}$ ($M_{\odot}$)} & {$\langle T\rangle_{\rm{NHI}}$ (K)} & {$\langle n_{\rm{gas}}\rangle_{\rm{NHI}}$ (cm$^{-3}$)} \\\hline \vspace{0.1cm}
 1.3$\times 10^{10}$& 1.0490$\times 10^{4}$ & 1.29$\times 10^{-2}$ \\
 1.0$\times 10^{10}$& 1.0486$\times 10^{4}$ & 1.30$\times 10^{-2}$ \\
 8.6$\times 10^{9}$& 1.0494$\times 10^{4}$ &  1.31$\times 10^{-2}$\\
 7.5$\times 10^{9}$& 1.0490$\times 10^{4}$ &  1.31$\times 10^{-2}$\\
 6.6$\times 10^{9}$& 1.0488$\times 10^{4}$ &  1.31$\times 10^{-2}$\\\hline
\end{tabular}
\label{tab:converg}
\end{table}

\section{Conclusions}
\label{sec:conclusion}

Metal-poor (${\rm [Z/H]<-2}$) damped Lyman alpha systems are thought to contain useful information
about the cold gas streams feeding galaxies at high redshift.
Utilizing the high-resolution, large-scale LAOZI cosmological simulations we, for the first time, 
investigate the physical nature of mpDLAs at $z=3$.
The major results found can be summarized as follows.
The majority of mpDLAs inhabit mostly regions $>20$ kpc from the host galaxy center on infalling cold gas streams originating from the intergalactic medium, 
with infall velocity of $\sim 100$ km/s, temperature of $\sim 10^{4}$ K and $\sim 1\%$ covering fraction within a radius $150$ kpc of high-z galaxies. 
They are relatively diffuse ($n_{\rm{gas}} \sim 10^{-2}$ cm$^{-3}$), 
Jeans quasi-stable and have very low star formation rate ($\dot{\Sigma} \le 10^{-4} \msun \rm{\ yr}^{-1} \rm{\ kpc}^{-2}$). 
They continue to migrate inward to the galaxy center while being mixed with high metallicity gas and stellar outflows
in the process, removing themselves from the metal-poor category and rendering 
the central ($\le 5$ kpc) regions of galaxies devoid of mpDLAs,
where, in sharp contrast, metal-rich DLAs have high a covering fraction.
All observables of the simulated mpDLAs are in excellent agreement with observations \citep[][]{2014Cooke},
except for the gas density, 
$n_{\rm{gas}} \sim 10^{-2}$ cm$^{-3}$, found in the simulation versus 
$n_{\rm{gas}} \sim 10^{-1}$ cm$^{-3}$
inferred from observations \citep[][]{2014Cooke}.
We tentatively attribute the cause of the discrepancy to some of the simplifications
made in the observational modeling,
including slab geometry and constant density, which are not borne out in our simulations.

\section*{Acknowledgments}
We would like to thank Dr. Ryan Cooke for useful discussion.
Analysis and visualization are in part made with the software program yt \citep[][]{2011Turk}.
Computing resources were in part provided by the NASA High-
End Computing (HEC) Program through the NASA Advanced
Supercomputing (NAS) Division at Ames Research Center.
This work is supported in part by grant NASA NNX11AI23G.
The simulation data are available from the author upon request.

\bibliographystyle{mn2e}
\bibliography{astro}

\appendix

\bsp

\label{lastpage}

\end{document}